# Subband adaptive filter trained by differential evolution for channel estimation

Lu Lu and Haiquan Zhao

**Abstract.** The normalized subband adaptive filter (NSAF) is widely accepted as a preeminent adaptive filtering algorithm because of its efficiency under the colored excitation. However, the convergence rate of NSAF is slow. To address this drawback, in this paper, a variant of the NSAF, called the differential evolution (DE)-NSAF (DE-NSAF), is proposed for channel estimation based on DE strategy. It is worth noticing that there are several papers concerning designing DE strategies for adaptive filter. But their signal models are still the single adaptive filter model rather than the fullband adaptive filter model considered in this paper. Thus, the problem considered in our work is quite different from those. The proposed DE-NSAF algorithm is based on real-valued manipulations and has fast convergence rate for searching the global solution of optimized weight vector. Moreover, a design step of new algorithm is given in detail. Simulation results demonstrate the improved performance of the proposed DE-NSAF algorithm in terms of the convergence rate.

## 1 Introduction

When the input signals are highly correlated, some stochastic gradient algorithms converge slowly. Therefore, the subband adaptive filters (SAF) with multiband structure, called the normalized SAF (NSAF) algorithm, was proposed to address this problem [2]. The main idea of the NSAF is to use the subband signals, normalized by their respective subband input variances, to update the weights of the fullband adaptive filter. This strategy leads to the decorrelating property of the NSAF algorithm. The estimation of parameters by using SAF has been proposed in several previous studies, including [1,3,4]. Particularly, Ni et al. proposed the variable regularization matrix NSAF (VRM-NSAF) algorithm via largest mean square deviation (MSD) decrease method [4]. Unfortunately, it has the high computational complexity, which may prohibit its practical applications. Abadi et al. proposed the SM-NSAF algorithm based on set-membership (SM) scheme [1]. Not only low-cost of this algorithm has, but also it achieves the fast convergence rate and small misadjustment. In these works and other similar references on the topic, the solutions typically rely on the use of a gradient descent method and less attention is paid to non-gradient solutions.

On the other hand, differential evolution (DE) algorithm is a simple powerful algorithm for real parameter optimization, which is based on the theory of swarm intelligence (SI) [8]. It has been shown to perform better than the genetic algorithm (GA) over several numerical benchmarks. Like GA, three fundamental processes drive the evolution of the DE algorithm, i.e., mutation, crossover, and selection. By using these mechanisms, the DE algorithm is able to explore different regions of the search space and previous knowledge about the fitness landscape. The DE algorithm has been employed in the diverse fields, such as engineering design [10], image processing [7] and scheduling problem [9], etc.

Typically, in past decades, the DE algorithm has been successfully implemented in the design of digital filters [5-6]. In the design of digital filter, the designer always compromised on one or the other of the design specifications. In contrast, the algorithm can provide the enhanced performance by using evolutionary methods (such as DE and GA algorithms). Since population-based stochastic search methods have proven to be effective in multidimensional environment, all the parameters of the filter can be effectively selected by adopting these algorithms.

In this work, the NSAF is designed based on the DE algorithm for accelerating the convergence speed. The proposed DE-NSAF algorithm combines benefits of the NSAF with multiband structure and the DE algorithm, and is a more efficient scheme for global optimization over continuous spaces. The weight vector of the SAF is adjusted by training the DE algorithm, making the proposed algorithm obtain an enhanced performance for channel identification. Our ma-

L. Lu • H. Zhao(✉)
School of Electrical Engineering, Southwest Jiaotong University, Chengdu, China
e-mail: lulu@my.swjtu.edu.cn, hqzhao_swjtu@126.com

jor contributions of this paper can be summed up in the following points: (1) to apply DE strategy to the NSAF algorithm, (2) a complete design steps of the proposed DE-NSAF algorithm is outlined for channel estimation, and (3) comparable results with other NSAF-based algorithms for channel estimation problem are presented.

## 2 Review of existing algorithms

Consider the desired response $\{d(n)\}$ arising from the model [2]:

$$d(n) = \mathbf{u}^T(n)\mathbf{w}_o + v(n) \tag{1}$$

where $\mathbf{w}_o = [w_{o,0}, w_{o,0}, ..., w_{o,M-1}]^T$ is the weight vector of the unknown channel, $\mathbf{u}(n) = [u(n), u(n-1), ..., u(n-M+1)]^T$ represents the input signal vector with length $M$, and $v(n)$ is the noise signal.

♦ NSAF algorithm

Lee and Gan proposed the NSAF algorithm to avoid aliasing components of the filter output terminal [2], where the input signal $\mathbf{u}(n)$ and the desired signal $d(n)$ are split by the analysis filter to form the subband signal $u_i(n)$ and $d_i(n)$, $i = 1, 2, ..., N-1$. The output signal $y_i(n)$ is obtained by $u_i(n)$ passing through the adaptive filter $\mathbf{w}(n)$. Then, the decimated filter output $y_{i,D}(k)$ at each subband can be written as

$$y_{i,D}(k) = y_i(kN) = \sum_{i=0}^{M-1} u_i(kN-m)w_m(k) = \mathbf{u}_i^T(k)\mathbf{w}(k) \tag{2}$$

where $\mathbf{w}(k) = [w_0(k), w_1(k), ..., w_{M-1}(k)]^T$, $\mathbf{u}_i(k) = [u_i(kN), u_i(kN-1), ..., u_i(kN-M+1)]^T$, $n$ and $k$ are used to index the original and decimated signals, respectively. The $i$th decimated subband error signal can be given by

$$e_{i,D}(k) = d_{i,D}(k) - y_{i,D}(k) \tag{3}$$

where $d_{i,D}(k)$ is the decimated subband desired signal. By using the principle of minimal disturbance, the adaptation of the weight vector of the original NSAF can be expressed as [2]:

$$\mathbf{w}(k+1) = \mathbf{w}(k) + \mu \sum_{i=0}^{N-1} \frac{\mathbf{u}_i(k)}{\delta + \|\mathbf{u}_i(k)\|^2} e_{i,D}(k) \tag{4}$$

where $\mu \in (0,2)$ is the step size, and $\delta$ is the regularized parameter.

♦ SM-NSAF algorithm

Introducing SM to the NSAF algorithm, the SM-NSAF algorithm is proposed [1], which has a lower computational burden and faster convergence rate as compared with the NSAF algorithm. The filter vector update equation of the SM-NSAF algorithm can be expressed as [1]:

$$\mathbf{w}(k+1) = \mathbf{w}(k) + \mu \sum_{i=0}^{N-1} \alpha_i(k) \frac{\mathbf{u}_i(k)}{\|\mathbf{u}_i(k)\|^2} e_{i,D}(k) \tag{5}$$

where

$$\alpha_i(k) = \begin{cases} 1 - \dfrac{\gamma}{|e_{i,D}(k)|}, & if\ |e_{i,D}(k)| > \gamma \\ 0, & otherwise \end{cases}, i = 0, 1, ..., N-1. \tag{6}$$

Note that $\gamma$ is the error bound which is related to the power of the system noise.

## 3 Proposed DE-NSAF algorithm

In this section, the DE algorithm is used to train the weight vector of the SAF $\mathbf{w}(n)$ for channel estimation. First, we define the cost function as

$$f = \sum_{i=0}^{N-1} e_{i,D}^2(k) = \sum_{i=0}^{N-1} \{d_{i,D}(k) - y_{i,D}(k)\}^2. \tag{7}$$

Following the definition of the cost function, an initial population consisting of many parameter vectors (candidate solutions) is then conducted.

Next, the DE for SAF begins with a population size of PS $M$-dimensional (i.e. the length of the SAF) parameter vectors representing the candidate solutions. Define $G = 0,1,...,G_{\max}$ as the subsequent generations in DE, and the $i$th vector of the population at the current generation is notation as

$$\vec{X}_{i,G} = \left[x_{1,i,G}, x_{2,i,G}, x_{3,i,G}, ..., x_{M,i,G}\right]. \tag{8}$$

Then, randomly generate an initial population from the interval [−1,1]. Every manipulated parameter vector will adapt three important mechanisms including mutation, crossover, and selection. Executing these operations one time is called a generation.

♦ Mutation

After initialization, DE employs the mutation operation to produce a mutant vector. Through mutation operation, DE creates a donor vector $\vec{V}_{i,G} = \left\{v_{1,i,G}, ..., v_{M,i,G}\right\}$ corresponding to each population member or target vector $\vec{X}_{i,G}$ in the current generation. The expression of differential mutation operation is given by [8]:

$$\vec{V}_{i,G} = \vec{X}_{r_1^i,G} + K \cdot \left(\vec{X}_{r_2^i,G} - \vec{X}_{r_3^i,G}\right). \tag{9}$$

The indices $r_1^i, r_2^i, r_3^i \in \{1,2,...,PS\}$ are mutually different indices randomly chosen from the current generation [1,2,...,PS], and $K \in (0,2)$ is a scaling factor that controls the amplification of the differential variation $\vec{X}_{r_2^i,G} - \vec{X}_{r_3^i,G}$.

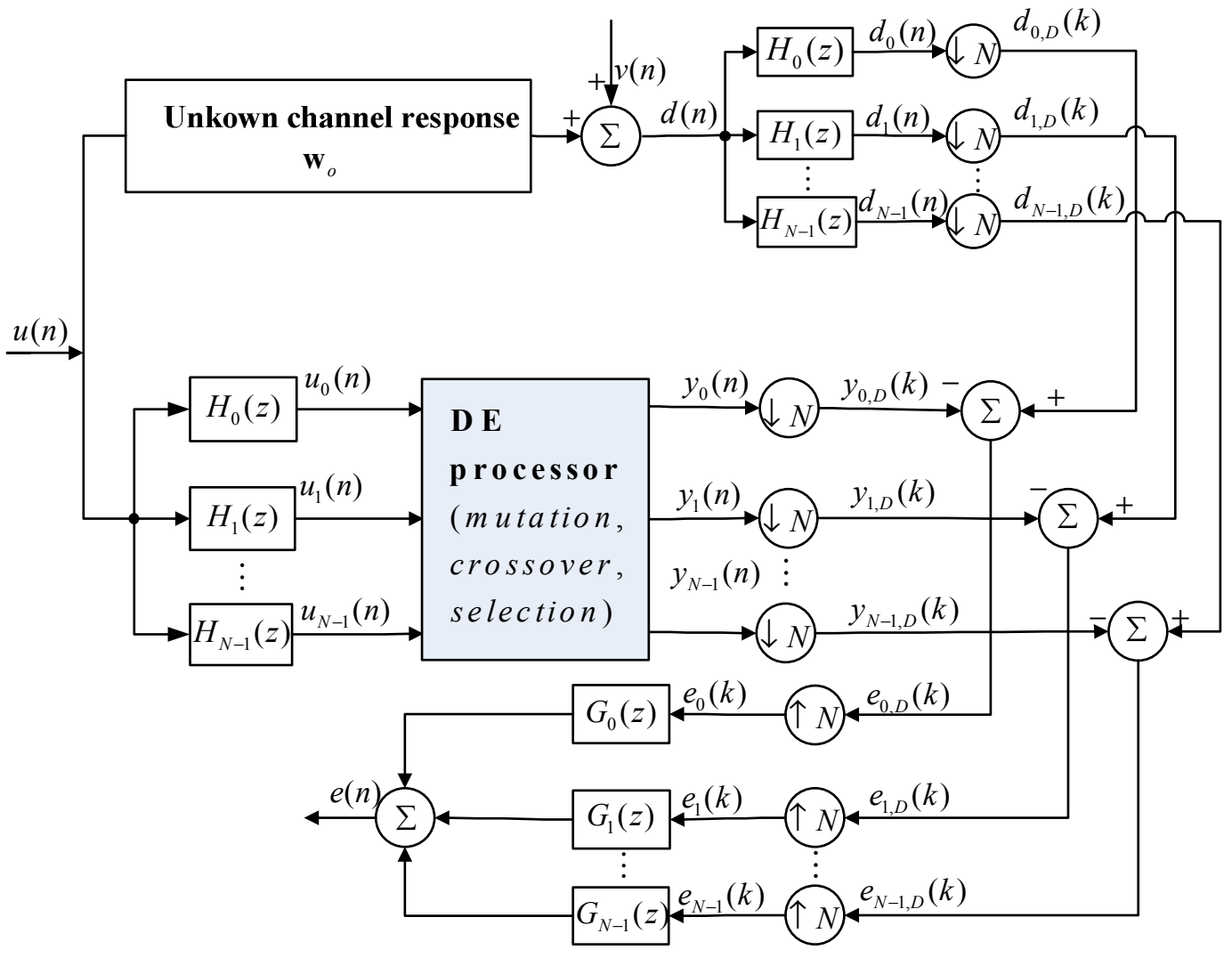

**Fig. 1** Block diagram of the DE-NSAF algorithm

♦ Crossover

The crossover operation is to interchange the components between mutant vector $\vec{V}_{i,G}$ and target vector $\vec{X}_{i,G}$ [8]. The DE algorithm can use two kinds of the crossover schemes—exponential and binomial. Throughout this paper, the binomial crossover operation is applied. The binary sequences $p_i$ is derived according to

$$p_i = \begin{cases} 1, & \beta_{ij} \le C_r \\ 0, & otherwise \end{cases} \tag{10}$$

where $\beta_{ij} \in (0,1)$ , and $C_r$ is the crossover probability range from (0,1). Finally, a trial vector $\vec{U}_{i,G} = \left[u_{1,i,G}, u_{2,i,G}, u_{3,i,G}, ...u_{M,i,G}\right]$ is obtained from

$$u_{j,i,G} = \begin{cases} v_{j,i,G}, & if\ p_i = 1 \\ x_{j,i,G}, & if\ p_i = 0. \end{cases} \tag{11}$$

**Table 1** Summary of the proposed DE-NSAF algorithm.

**%start DE-NSAF algorithm**
**for** $k$=1 **to iteration number**
**Select the parameter** $N$, $M$ **of NSAF algorithm**

$$d(n)\xrightarrow{H_{N-1}(z)}d_i(n)\xrightarrow{\text{discretization}}d_{i,D}(k)$$

$$u(n)\xrightarrow{H_{N-1}(z)}u_i(n)\xrightarrow{\text{discretization}}u_{i,D}(k)$$

**%start DE processor**
**%*Initialization of DE algorithm*%**
**Select the parameter of DE algorithm** $C_r$, $K$, $PS$, **and** $G_{\max}$,
**Randomly generated an initial population from** [-1,1]
**for** $G$=1 **to** $G_{max}$
**%*Mutation operation*%**
**for** $i$=1 to $PS$
**Generate a donor vector** $\vec{V}_{i,G}=\{v_{1,i,G},\ldots,v_{M,i,G}\}$
**corresponding to the** $i$th **target vector** $\vec{X}_{i,G}$ **via**
**the different mutation in (11).**
**end for**

**%*Crossover operation*%**
**for** $i$=1 to $PS$
**Generate a trial vector** $\vec{U}_{i,G}=\left[u_{1,i,G},u_{2,i,G},u_{3,i,G},\ldots u_{M,i,G}\right]$
**for the** $i$th **target vector** $\vec{X}_{i,G}$ **through binomial**
**rossover operation in (12) and (13).**
**end for**

**%*Selection operation*%**
**for** $i$=1 to $PS$
**Compute the cost function of** $\vec{U}_{i,G}$ **and** $\vec{X}_{i,G}$
**if** $f(\vec{U}_{i,G})<f(\vec{X}_{i,G})$
**Then** $\vec{X}_{i,G+1}=\vec{U}_{i,G}$
**else**
$\vec{X}_{i,G+1}=\vec{X}_{i,G}$
**end if**
**end for**
**return the global best solution** $\mathbf{w}(k)=\vec{X}_{i,G+1}$
**end for**
**end DE processor**
**compute the output of the filter** $y_{i,D}(k)=\mathbf{u}_i^T(k)\mathbf{w}(k)$
**compute the error signal of the filter** $e_{i,D}(k)=d_{i,D}(k)-y_{i,D}(k)$
**end DE-NSAF algorithm**

From (11), it concludes that the trail vector $\vec{U}_{i,G}$ is a result of sufficiently exchanging vectors $\vec{V}_{i,G}$ and $\vec{X}_{i,G}$. The crossover operation described in (10)-(11) is basically a discrete recombination.

♦ Selection

The selection operation is to determine whether the target or the trial vector survives to the next generation. The cost function of the trial vector $\vec{U}_{i,G}$ and target vector $\vec{X}_{i,G}$ are needed to respectively evaluate. At $G$=$G$+1, if $f\left(\vec{U}_{i,G}\right)<f\left(\vec{X}_{i,G}\right)$, then the target vector is replaced by the trail vector, otherwise, the target vector survives in the population, i.e.,

$$\vec{X}_{i,G+1} = \begin{cases} \vec{U}_{i,G}, & if\left(\vec{U}_{i,G}\right) < f\left(\vec{X}_{i,G}\right) \\ \vec{X}_{i,G}, & otherwise \end{cases} . \tag{12}$$

In the proposed algorithm, $G=G_{max}$ is adopted as a termination condition. Using above three steps of the DE algorithm (DE processor), the weight vector of the subband adaptive filter is adapted. The block diagram of the proposed algorithm is shown in Fig. 1.

In summary, the design step for DE-NSAF algorithm can be outlined in Table 1. The objective of the DE-NSAF algorithm is to minimize the mean square error of the subband adaptive filter.

**Input Data**: Parameters *N*, *M* of the subband adaptive filter, population size *PS*, crossover probability $C_r$, scaling factors *K*, and number of generations $G_{max}$.

**Output Data**: Fullband adaptive weight vector $\mathbf{w}(k)$.

## 4 Simulations

To demonstrate the performance of the proposed method, the channel identification example is illustrated in this section. All the simulation results are obtained by averaging over 200 independent trials. The unknown channel has 32 taps, selected at random. The input signal $u(n)$ is a fourth order autoregressive (AR(4)) signal expressed by [1]

$$\begin{aligned} u(n) &= 0.6617u(n-1) + 0.3402u(n-2) \\ &+ 0.5235u(n-3) - 0.8703u(n-4) + \xi(n) \end{aligned} \tag{13}$$

where $\xi(n)$ is a zero mean white Gaussian noise. In all simulations, the fixed parameters are given by scaling factors $K$=0.5, number of generations $G_{max}$=3000, and the number of subband $N$=4. The signal-to-noise ratio (SNR) is chosen to be 20dB. Firstly, we let the population size of the DE vary within [10, 50], in order to test the performance of algorithm under different population sizes. Fig. 2 shows the performance of the algorithm under different population sizes. To demonstrate the effect of the population sizes clearly, the number of generations is truncated to 100 points. For $PS$>20, the convergence rate is slow. Moreover, the small population size ($PS$=10) leads to poor convergence property. It turns out that the best option is $PS$=20. Fig. 3 illustrates the effect of $C_r$ on DE-NSAF under $PS$=20. As can be seen, the large value of $C_r$ has high mutation rate and fast convergence rate, but tends to obtain local optimum solution. Consequently, the best option is $C_r$=0.8 under overall consideration. Fig. 4 shows a mean square error (MSE) comparison of the NSAF, SM-NSAF and the proposed algorithms. In SM-NSAF, $\gamma$ is set to $\sqrt{5\sigma_v^2}$ [1], where $\sigma_v^2$ is the noise power. It is observed that the proposed algorithm achieves faster convergence rate than the NSAF and SM-NSAF, the MSE achieved by the DE-NSAF filter is −19dB, whereas the SM-NSAF and NSAF ($\mu$=0.1) achieve approximately −17dB and −18.5dB, respectively. Furthermore, the average MSEs of the algorithms in steady-state are shown in Table 2. Again, the proposed DE-NSAF algorithm has a comparable misadjustment to other algorithms in steady-state.

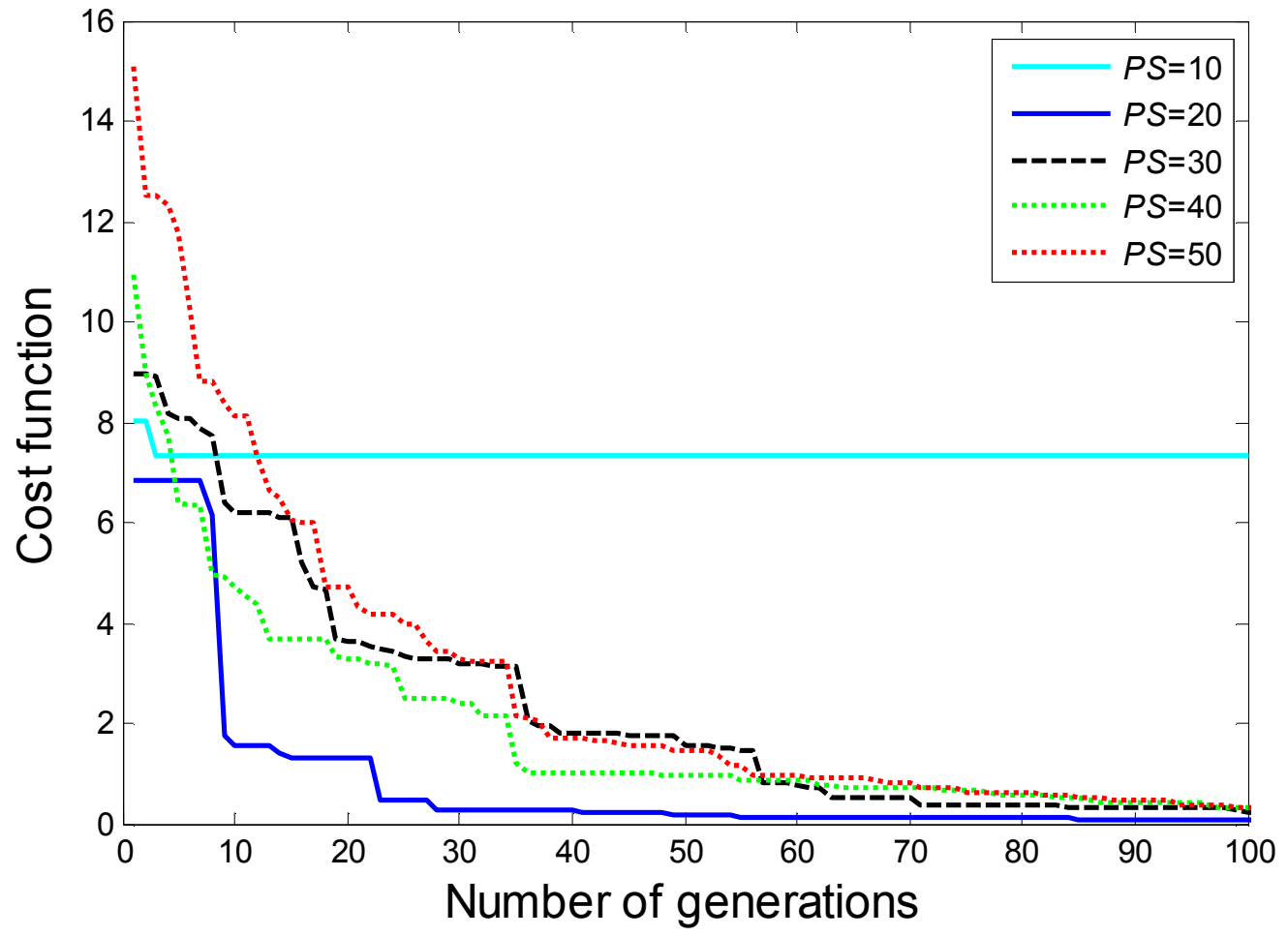

**Fig. 2** Effect of the Population Size (*PS*=10~50).

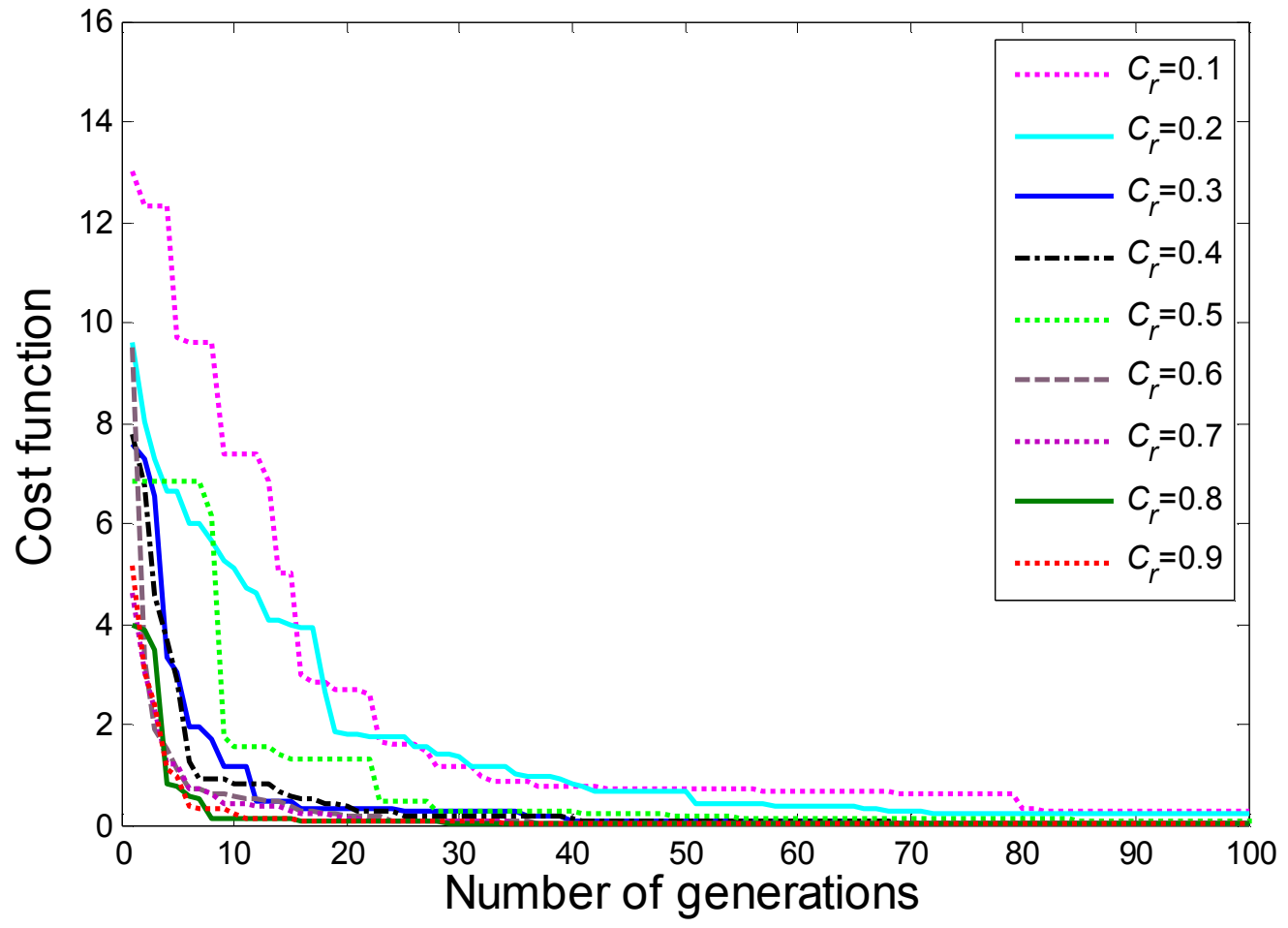

**Fig.3** Effect of the $C_r$ (*PS* is fixed at 20).

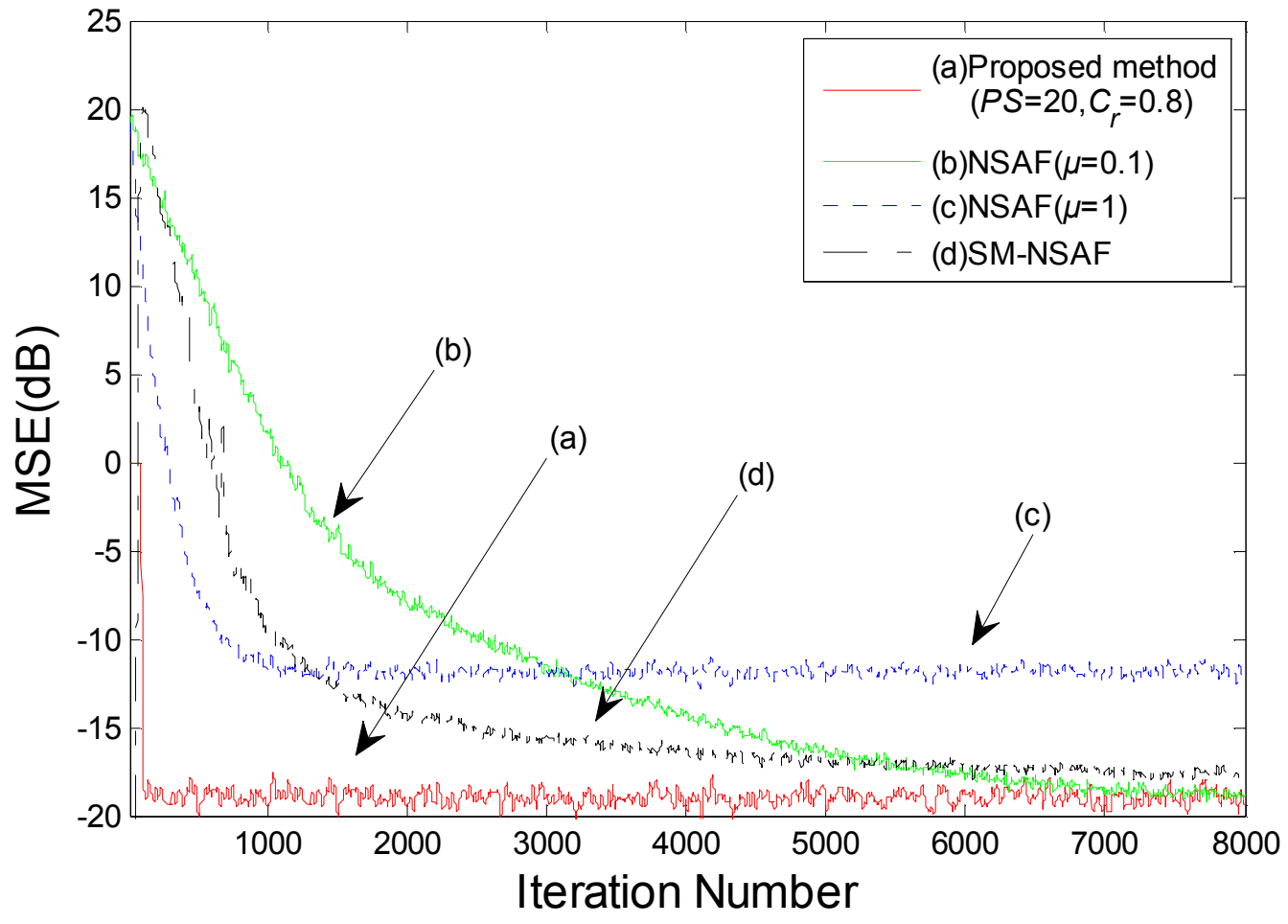

**Fig. 4** Learning curves of proposed method in SNR=20dB, *N*=4 (Input Gaussian AR(4)).

**Table 2** Comparison of average MSE for algorithms.

| Algorithms | Average MSE |
|---|---|
| **NSAF**($\mu$=0.1, $N$=4) | 0.016012±0.0021955 (5000~end) |
| **NSAF**($\mu$=1, $N$=4) | 0.065374±0.0042192 (1000~end) |
| **SM-NSAF** ($N$=4) | 0.019568±0.0018551 (4000~end) |
| **DE-NSAF** ($N$=4) | 0.012755±0.0011735 (100~end) |

# 5 Conclusions

This paper has successfully applied the DE algorithm into the NSAF for channel estimation. The weight vectors are adjusted according to three evolutionary mechanisms of the DE. Due to the property of real-valued manipulations, the proposed DE-NSAF algorithm is rather effective in solving the problem to others. Simulation results in the context of channel estimation demonstrate that the proposed method achieves fast convergence property as compared with the NSAF and SM-NSAF algorithms.

## Acknowledgments

This work was supported in part by National Natural Science Foundation of China (Grants: 61571374, 61271340, 61433011).

## References

1. M.S.E. Abadi, J.H. Husøy, Selective partial update and set-membership subband adaptive filters. Signal Processing 88 (2008) 2463–71.
2. K.A. Lee, W.S. Gan, Improving convergence of the NLMS algorithm using constrained subband updates. IEEE Signal Processing Letters 11 (2004) 736–9.
3. L. Lu, H. Zhao, C. Chen, A normalized subband adaptive filter under minimum error entropy criterion. Signal, Image and Video Processing 10 (2016) 1097–1103.
4. J. Ni, F. Li, A variable regularization matrix normalized subband adaptive filter. IEEE Signal Processing Letters 16 (2009) 105–8.
5. K.S. Reddy, S.K. Sahoo, An approach for FIR filter coefficient optimization using differential evolution algorithm, International Journal of Electronics and Communications 69 (2015) 101–8.
6. N. Salvatore, A. Caponio, F. Neri, S. Stasi, et. al. Optimization of delayed-state Kalman-filter-based algorithm via differential evolution for sensorless control of induction motors. IEEE Transactions on Industrial Electronics 57 (2010) 385–94.
7. S. Sarkar, S. Das, S.S Chaudhuri, A multilevel color image thresholding scheme based on minimum cross entropy and differential evolution. Pattern Recognition Letters 54 (2015) 27–35.
8. R. Storn, K.V. Price, and J. Lampinen, Differential Evolution–A Practical Approach to Global Optimization. Berlin, Germany: Springer-Verlag, 2005.
9. A. Trivedia, D. Srinivasana, S. Biswasc, T. Reindl, Hybridizing genetic algorithm with differential evolution for solving the unit commitment scheduling problem. Swarm and Evolutionary Computation 23 (2015) 50–64.
10. J.T. Tsai, Improved differential evolution algorithm for nonlinear programming and engineering design problems. Neurocomputing 148 (2015) 628–40.